# The design and the performance of an ultrahigh vacuum $^3$He fridge-based scanning tunneling microscope with a double deck sample stage for *in-situ* tip treatment

Syu-You Guan,[a,b,1] Hsien-Shun Liao,[a,c] Bo-Jing Juang,[a] Shu-Cheng Chin,[a] Tien-Ming Chuang,[a,2] and Chia-Seng Chang [a,b,3]

[a] *Institude of Physics, Academia Sinica, Nankang, Taipei 11529, Taiwan*

[b] *Department of Physics, National Taiwan University, Taipei 10617, Taiwan*

[c] *Department of Mechanical Engineering, National Taiwan University, Taipei 10617, Taiwan*

**Abstract**

Scanning tunneling microscope (STM) is a powerful tool for studying the structural and electronic properties of materials at the atomic scale. The combination of low temperature and high magnetic field for STM and related spectroscopy techniques allows us to investigate the novel physical properties of materials at these extreme conditions with high energy resolution. Here, we present the construction and the performance of an ultrahigh vacuum $^3$He fridge-based STM system with a 7 Tesla superconducting magnet. It features a double deck sample stage on the STM head so we can clean the tip by field emission or prepare a spin-polarized tip *in situ* without removing the sample from the STM. It is also capable of *in situ* sample and tip exchange and preparation. The energy resolution of scanning tunneling spectroscopy at T = 310 mK is determined to be 400 mK by measuring the superconducting gap with a niobium tip on a gold surface. We demonstrate the performance of this STM system by imaging the bicollinear magnetic order of $Fe_{1+x}Te$ at T = 5 K

**Keywords**

Spin polarized scanning tunneling microscopy
Superconducting gap
Magnetic structure

___________________________

[1.] Electronic mail: r95222038@ntu.edu.tw.

[2.] Electronic mail: chuangtm@gate.sinica.edu.tw.

[3.] Electronic mail: jasonc@phys.sinica.edu.tw.

## 1. Introduction

The invention of scanning tunneling microscope (STM) [1] enables the direct visualization of the surface of condensed matter systems at the atomic scale. More importantly, scanning tunneling spectroscopy (STS) [2] provides the key information for understanding the physical properties of the materials by imaging the spatial dependence of the local density of states at different energies; whereas, Fourier transform-scanning tunneling spectroscopy (FT-STS) has become a powerful technique to investigate the electronic structure in momentum space by imaging quasiparticle scattering interference (QPI) [3]. Furthermore, the development of spin-polarized STM [4] allows us to directly image the atomic spin structures of magnetic materials in real space. Thus, STM has become an indispensable tool for the research of condensed matter systems. In addition, strongly correlated electron systems such as high Tc superconductors and heavy fermion superconductors often exhibit a complex phase diagram [5]. To understand these unique phases of matters, it is necessary to perform the STM and STS measurements at low temperature in the presence of a magnetic field with high energy resolution. Consequently, this has been the motivation for the development of STM systems toward millikelvin temperature with a high magnetic field [6-24]. However, building such a STM system remains challenging because the STM operation in the combination of ultra-low temperature, high magnetic field and ultra-high vacuum (UHV) environments poses stringent constraints for the STM design and optimization. Moreover, the requirements for the *in situ* exchange of the STM tip and the sample, the sample preparation, the cryogenic cleave stage for single crystals and the installation of other surface analytical tools further complicates the system construction.

In this paper, we describe the design and the performance of our homemade UHV $^3$He fridge-based STM ($^3$He - STM) with a 7 Tesla superconducting magnet along the z-axis. A rigid and compact Pan-style STM head [6] is mounted on a single shot $^3$He fridge cryostat that can cool down the sample below 300mK for approximately 24 hours. The system can also be operated at T = 5 K for ~2 days before having to refill the liquid helium. We design a double deck sample stage on the STM head so that the sample can be preserved inside the STM head while the STM tip can be cleaned on the gold surface by field emission or a spin-polarized tip can be prepared on a magnetic sample *in situ*. The STM and $^3$He fridge are installed in a multi-chamber UHV system, which allows us to prepare and exchange the samples as well as the tip *in situ*. To calibrate the electron temperature, we fit the superconducting gap measured by using a Nb tip on Au surface, from which we determine the electron temperature is T ~ 400 mK for STS. We demonstrate the performance of our STM by imaging the surface and bicollinear magnetic structure of $Fe_{1+x}Te$ single crystal with a normal tungsten and spin polarized tip at T = 5 K.

## 2. Material and methods

*2.1. System Overview*

The UHV $^3$He - STM system consists of three UHV chambers: the STM chamber, the preparation chamber and the load-lock chamber (Fig. 1) which allows us to transfer and prepare the tip and the samples between different chambers within UHV. After baking at 90°C and pumping by a turbo pump for three days, the system vacuum can reach about $1\times10^{-10}$ Torr by an ion pump. The UHV environment is subsequently maintained by an ion pump and by the cold surface of the dewar. For the STM chamber, a fully UHV compatible, 12-liters helium dewar with a 7-Tesla split-coil superconducting magnet is installed inside. The holding time is about 48 hours at T = 5 K. The STM head is located at the center of the magnet bore which has the homogeneity of ±0.1% in a 1 cm diameter spherical volume. The tip and the samples can be accessed into STM through one side of the magnet bore. A molecular beam epitaxy (MBE) gun (Omicron model EFM-3) is equipped on the other side of the magnet bore such that we can perform the atomic deposition onto the sample surface when the sample is still in the STM head. A sample and tip parking stage is also installed in the STM chamber for storing multiple samples and tips in UHV.

The preparation chamber consists of an ion sputter gun, an electron-bombardment heating stage and two MBE guns for the sample cleaning, the sample deposition and the tip preparation. One electron-bombardment stage is installed for the heat treatment of the sample and tip. A low energy electron diffraction (Omicron SPA-LEED) is installed for quick examination of the surface structure. Additional sample and tip parking stage is installed in the load-lock chamber, which allows us to exchange samples and tips in vacuum and thus, greatly reduce the turnaround time.

To isolate the vibration noise, the entire UHV STM system is supported on an optical table with four pneumatic springs (Newport model S-2000A). We install the ion pumps and additional ~600 kg of lead bricks at the bottom of the optical table to lower the center of mass of the system and improve the stability of vibration isolation system.

## 2.2. $^3$He Cryostat

The direct way to cool down the sample into the millikelvin region is to evacuate the liquid $^3$He. Typically a $^4$He pot or Joule-Thomson fridge is required to condense $^3$He gas into liquid first [25]. However, pre-cooling the $^4$He liquid to superfluid state is necessary to minimize the noise. To avoid this issue, we use a single shot $^4$He pot to condense $^3$He gas [26] as illustrated in Fig.2 (a). Two stainless steel gas containers are filled with $^4$He (8 liters in STP) and $^3$He (4 liters in STP), respectively. The absorption or desorption of Helium gas is controlled by the temperature of two individual charcoal pumps. The thermal links between the charcoal pumps and the helium dewar are connected or disconnected by turning on or off the two gas-gap heat switches (GGHS) [27], respectively. The STM head is mounted to the bottom of the $^3$He pot. The GGHS and the STM head are enclosed within a copper radiation shield that is thermally anchored to the 4K plate. A window is designed on the radiation shield and can be opened by a wobble stick so that we can access the STM head for the tip/sample exchange or the atomic deposition. Neither mechanical pumping system nor moving parts are used in this design, thus running $^3$He fridge does not generate additional vibration.

The procedure of operating the $^3$He fridge while the system is at liquid helium temperature is as the following. To condense $^3$He gas into liquid, we first pump the liquid $^4$He in the pot by turning on the GGHS between the charcoal for the $^4$He pot and the helium bath. Running the $^4$He pot alone can cool down the fridge and the STM head to T ~ 2.0K for ~20 minutes until the $^4$He pot is empty. Then, the GGHS between the charcoal for the $^3$He pot and the helium bath is turned on to pump the liquid $^3$He, which lowers the STM head to T ~ 280 mK with the holding time about 6 hours. We can improve the $^3$He fridge efficiency if longer holding time is desired. We first pump the helium dewar for ~4 hour to cool down the liquid $^4$He and $^3$He in the pots to T ~ 2.0K and then we pump the $^4$He pot to reach T ~ 1.2 K. This procedure allows us to reach the base temperature below 300 mK and prolong the holding time to 24 hours (additional 10 L of liquid helium is evaporated and the dewar needs to re-filled) while it still maintains the quiet operation of our pot design.

*2.3. STM Head*

Because the vibration transfer function is decided by the square of the quotient of the resonance frequence of damping system and STM [28], a rigid and compact STM head with high resonance frequency is important to suppress the low-frequency vibration noise. To achieve this, we adopted the Pan-type STM design [6] (Fig. 2 (b)). Our STM body is made of titanium because of the similar thermal expansion coefficient with all ceramic parts, and its excellent mechanical strength and its great machinability. The coarse approach or the walker consists of a triangular sapphire prism which holds the piezo-tube scanner in the center of STM body and six shear piezo stacks (PI, PIC-255) which provide the stick-slip motion. In order to exchange the tip and sample *in situ*, the traveling distance of the walker is designed to be 10 mm. The piezo scanner is a piezoelectric tube (EBL Product, PZT-8, 0.3"long, 1/8" OD and 0.015" wall thickness) with four external gold-plated electrodes for raster scanning and one internal electrode for feedback along the z-axis, which yields the scanning range about 1 μm$^2$ with ± 220 V at T = 250 mK. In order to keep the rigidity of the piezo scanner assembly, a tantalum tip holder is designed to be as light as possible. A beryllium copper wire is carefully bent and then soldered on the tip holder receptacle on the piezo scanner as a spring for clamping the tip holder. The sample is glued by silver epoxy (EPO-TEK H20E) or fixed by spring plate on the sample holder before being inserted to the double deck sample stage (the detail will be described in the next section) for measurements.

*2.4. In-situ tip exchange and treatment*

The transfer rod for exchanging the tips and the samples needs to pass through the 2 cm wide side bore from STM without introducing too much heat from room temperature. Therefore, a stainless steel tube (6 mm outer diameter, 40 cm long, with 0.6mm thickness) with a bayonet connector at the end is used to grab the tip and the sample holder (Fig. 3(a)). Turning the grabber clockwise or counterclockwise can lock or release the holders, respectively. The tip exchange mechanism is shown in Fig. 3 (b). In order to take out the tip, we first introduce an empty loader of tip holder into the STM chamber. Next, the tip and the scanner tube are moved up by the walker to the exchange position, and then the loader is pushed forward to catch the tip holder into its groove. Optical access through the openings on the STM body, the magnet bore and the UHV viewports allows us to make fine adjustment during the tip exchange. Last, the scanning tube is retracted by the walker to remove the tip holder from the tip receptor on the scanner tube. The steps in reversed order are used to insert the tip holder to the piezo scanner.

Field emission on the clean metal is a convenient way to clean the tungsten tip [28]. It has been recently reported several sophisticated solutions to perform field emission *in-situ* STM without removing the sample. One method is to mount both the sample under investigation and the gold on a two-sample holder. By using a linear walker stage to precisely offset these two samples over the tip, the tip treatment can be realized and then the measurement can resume on the original sample surface [24]. The other method is to have the gold sample rotate onto or away from the tip position by a rotating piezo walker [29]. However, adding a complex walker stage to STM head may decrease the resonance frequency of STM head. To avoid this issue, we design a simple double deck sample stage for inserting two sample holders in the STM head (Fig. 2(b)). When we need to clean the tip, we can insert the Au (111) sample into the bottom sample stage. After the tip approached on the Au (111) surface, we apply the bias voltage on the tip up to several hundred dc volts to perform the field emission. Moreover, the process can be applied to prepare a spin-polarized tip by inserting a magnetic sample into the bottom sample holder. The spin-polarized tip can be prepared either by picking up magnetic atoms from the sample surface [30, 31] or by field emission on the magnetic sample surface [32]. The same magnetic sample can be used as a calibration sample after the tip preparation. Such a design gives us a simple and convenient way to clean the tip without adding addition walker stage as well as to prepare the spin-polarized tip without installing additional evaporator for coating the tip. Also, the sample under investigation can be preserved in the STM head so the comparison can be made on the same field of view (FOV) with a spin-polarized tip and a normal metal tip.

*2.5. Sample cleaving for single crystal samples*

To obtain the clean surface and to prevent the possible surface reconstruction, it is required to cleave the layered single crystal sample *in-situ* at low temperature. First, we glue an magnetic iron cleaving rod onto the sample surface and slowly cooled down the sample by radiant cooling in the STM chamber before inserting into the STM. Then, a magnet mounted on the cleaver which is attached on the transfer rod (Fig. 3(c)) is pushed to collide onto the cleaving rod on the sample. At the same time, the cleaving rod is attracted onto the cleaver which can be removed from the chamber. By this method, the sample is cleaved in UHV at $T < 10$ K, and we can determine the outcome of cleavage by examing the residue sample on the cleaving rod.

*2.6 Electronics and wiring*

Fifty electrical leads made by Constantan are used to connect the temperature sensors, heaters, piezo tube and piezo stacks. To eliminate the cross-talk from the high voltage signals, custom-made superconducting coaxial cables (Cooner Wires, NbTi conductor with CuNi plating and phosphor bronze shield) are used for the bias and the tunneling current. All wires are made of the materials with small thermal conductivity and are carefully thermal-anchored to the $^4$He pot and the 4 K plate to prevent excessive heat from leaking into the sub-Kelvin region. Radio frequency (RF) noise is detrimental to the electron temperature of STM, diminishing the energy resolution for STS. To attenuate the RF noise, the system is built in the RF shielding room. Furthermore, all signals pass the low pass $\pi$ filters before going to the STM chamber except the wires for tunneling current and the walker piezo stacks. The walker piezo stacks are connected to the ground after coarse approach. To eliminate the ground loops, all connections are carefully routed to single isolated ground and only the necessary electronic instruments (such as STM controller, thermometer and liquid level controller) are connected during the measurement. A current amplifier with a gain of $10^9$ (NF Corp, model SA-606F2) and the bandwidth of 3 kHz is used for tunneling current. Nanonis controller is used for STM control and data acquisition.

## 3. Results

*3.1. Current noise spectrum*

The current noise spectrum of STM at T = 300 mK is showed in Fig. 4(a). When the tip retreats from the sample surface, no particular noise frequencies are detected. We then take the current noise spectrum with a W tip on Au (111) surface and z-feedback turned off. The noise level is about 10 fA/ √Hz from DC to 1kHz, which is 1.5 times to the current noise density of current amplifier. The 28Hz noise is due to the nature frequency of the optical table and its supporting frame, which is verified by geophone measurement. The 60 and 180 Hz noise due to the pickup of the power line noise and its harmonics are always present. Although they have been minimized in our STM, it is difficult to completely eliminate them. We then take the current noise spectrum with a W tip on Au (111) surface and z-feedback turned off.

*3.2. Topographic and magnetic imaging*

We further demonstrate the STM performance by measuring a $Fe_{1+x}Te$ single crystal, which exhibits a unique bicollinear stripe magnetic structure at the low temperature. Fig. 4(c) shows the topography of the cleaved surface of $Fe_{1+x}Te$ single crystal obtained by using a normal tungsten tip. The Te lattice and excessive Fe clusters is clearly resolved with atomic resolution at T = 5 K. The Fourier transform of Fig. 4(c) only shows the Bragg peaks (inset). For SP-STM measurement, we adopt the same scheme as in Ref. [30, 31], which does not require a sophisticated preparation for the tip and the calibration sample. The spin-polarized tip is prepared by picking up the magnetic clusters on the surface of $Fe_{1+x}Te$ in Fig. 4(c) and then is calibrated on the same surface by imaging the bicollinear stripe magnetic structure of $Fe_{1+x}Te$ (the process is shown in Appendix). Fig 4(d) shows a topographic image taken at a location with three magnetic domains after the tip preparation. The Fourier analysis on the two larger domains (domain A and B) reveals the additional peaks due to the antiferromagnetic order at (±0.5, 0) along a-axis. The charge density wave (CDW) order at (±1, 0) along the a-axis is superimposed with the Bragg peaks, which show the stronger amplitude than that of the Bragg peaks at (0, ±1) along the b-axis. Furthermore, all characteristic peaks for the antiferromagnetic order, the CDW order and the Bragg peaks rotate through 90° across the domain boundary. These results are consistent with previous SP-STM studies on $Fe_{1+y}Te$ single crystals [30, 31, 33] and mono/bi-layered FeTe on $B_2Te_3$ [34], which indicates the magnetic tip has significant in-plane magnetic moment for resolving the in-plane magnetic structure in $Fe_{1+x}Te$. In contrast, an early STM study by using non-magnetic Pt/Ir tips on epitaxial FeTe films grown on 6H-SiC reported a charge order induced modulation with the peaks at (±0.5, 0) along a-axis [35]. In order to further examine the magnetic propertes of the tip, we measure $Fe_{1+x}Te$ under the magnetic field larger than the switching field of the ferromagnetic tip [31]. Topographic images taken in the domain B at -3T and +3T in the same FOV are shown in the top and bottom of Fig. 4(e), respectively. The magnetic contrast of the stripes in these images remains the same indicates the tip is antiferromagnetic probably because the tip picks up a cluster of $Fe_{1+y}Te$ at the apex as illustrated in Fig. 4(f) [31].

### *3.3. Superconducting gap spectrum*

To calibrate the energy resolution of our STS measurement at the base temperature, we first use the superconducting Nb tip to perform the STS measurement on the Au (111) surface. The tunneling spectrum in Fig. 4(b) shows the superconducting gap of Nb. We then fit the measured superconducting spectrum with Dynes equation [36] and obtain the Δ is 1.4 meV and the $T_{eff}$, the effective electron temperature is 400 mK, which is very close to the sample temperature at T= 310 mK which indicates our optimization of STM performance at 310mK works very well.

## 4. Conclusions

We have presented the design and the performance of a UHV $^3$He fridge – based STM system, which features a unique double deck sample stage on the STM head. Our design provides a simple method to *in-situ* prepare a spin-polarized tip or to clean the tip inside the STM head without exchanging the sample.

The multiple samples and tips storage allows us to perform in situ exchange, which greatly reduces the turn-around time. The noise level of tunneling current is 100 fA/√Hz and the effective temperature of the STS is determined to be ~ 400 mK. Our results demonstrate this UHV STM system is a great tool for the research of surface science and novel quantum materials in sub-Kelvin temperature.

**Acknowledgment**

We thank Shih-Hsin Chang, Pin-Jui Hsu, Germar Hoffmann, Peter Wahl, Tetsuo Hanaguri and J. C. Séamus Davis for helpful discussion. We acknowledge Raman Sankar and Fangcheng Chou for providing us the single crystals. The machining of many precision parts for the instrument was done in the machine shop of Institute of Physics, Academia Sinica, and their skill and effort are highly appreciated. This work is supported by Academia Sinica and Ministry of Science and Technology, Taiwan. S.Y. Guan acknowledges the support from Academia Sinica Postdoctoral Fellowship. T.M. Chuang is grateful for the support of Golden Jade Fellowship from Kenda Foundation.

**Figures**

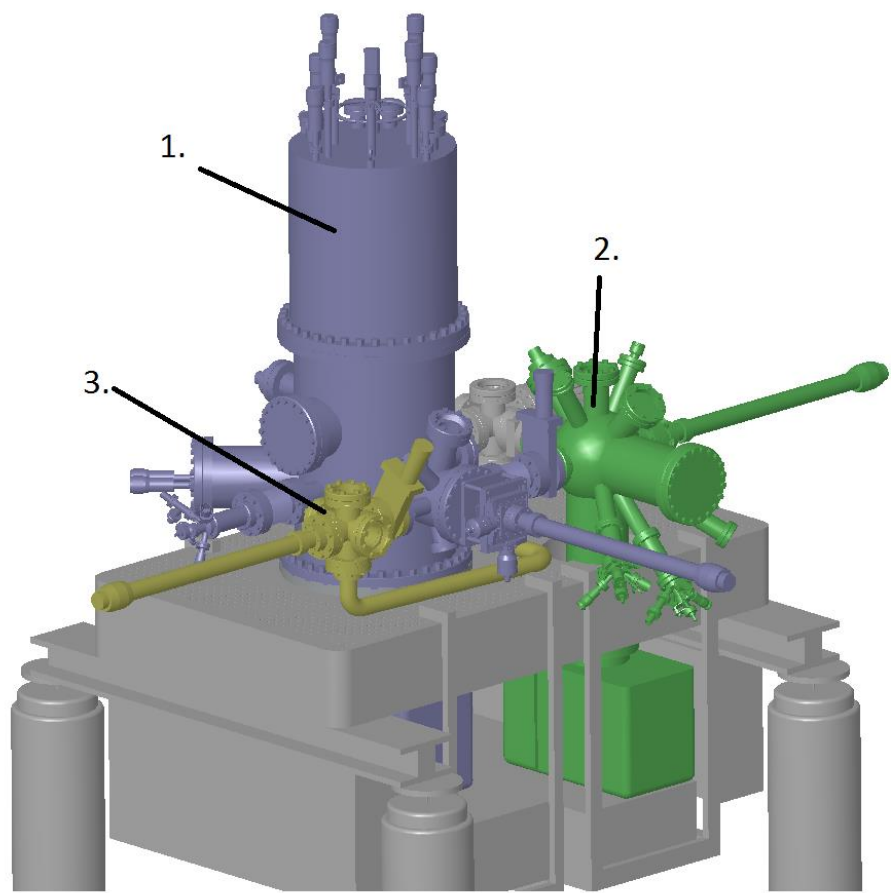

**Fig. 1.** The front view of the system. 1. STM chamber (purple). 2. Preparation chamber (green). 3. Load-lock chamber (yellow).

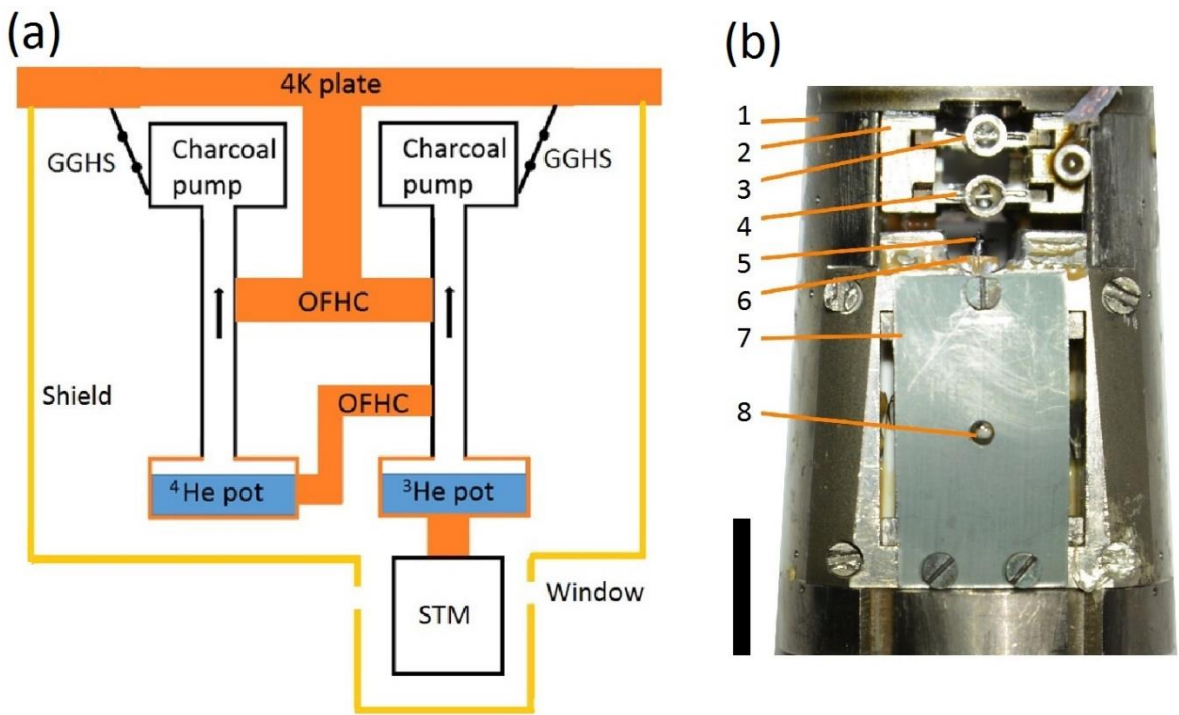

**Fig. 2.** (a) The scheme of the $^3$He cryostat. The $^4$He and $^3$He pot are made by oxygen free high conductivity copper for good thermal conductivity. On the other hand, the gas tubes are made of stainless steel to reduce the heat leak. The window on the radiation shield (yellow) can be opened or closed by wobble stick. (b) The photo of the front view of the STM head. (1) The STM main body. (2) The sample holder receptor. (3) The top sample is for the sample under investigation. . (4) The bottom sample is used for the tip treatment, field emission or calibration. (5) The tip holder. (6) Tip holder receptor glued on the piezo scanner. (7) The spring plate. (8) Sapphire ball for exerting force in a single point. The black scale bar represents 10mm.

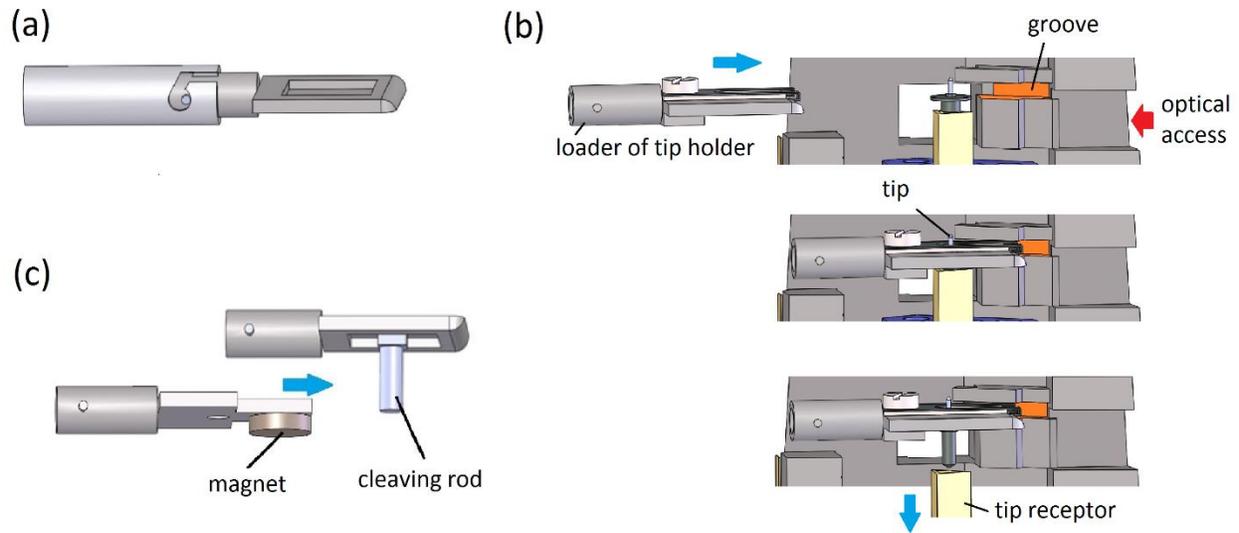

**Fig. 3. (a)** The grabber and the sample holder. **(b)** The section view of STM head with the schematic process of removing the tip holder from the piezo scanner. The blue arrows indicate the moving direction of each part. The grabber is attached to an UHV manipulator and the piezo tube is moved by the walker. The groove on the STM body (orange) is used to keep the loader at the exchange positon so that the manipulator will not be bent by the torque. The entire process can be performed with optical access through the opening on the STM body, the side bore of the magnet and the viewports on the chamber. **(c)** The cleaving mechanism shows the cleaver with a magnet collides with the cleaving rod which is glued on the sample. The blue arrow indicates the moving direction of the cleaver.

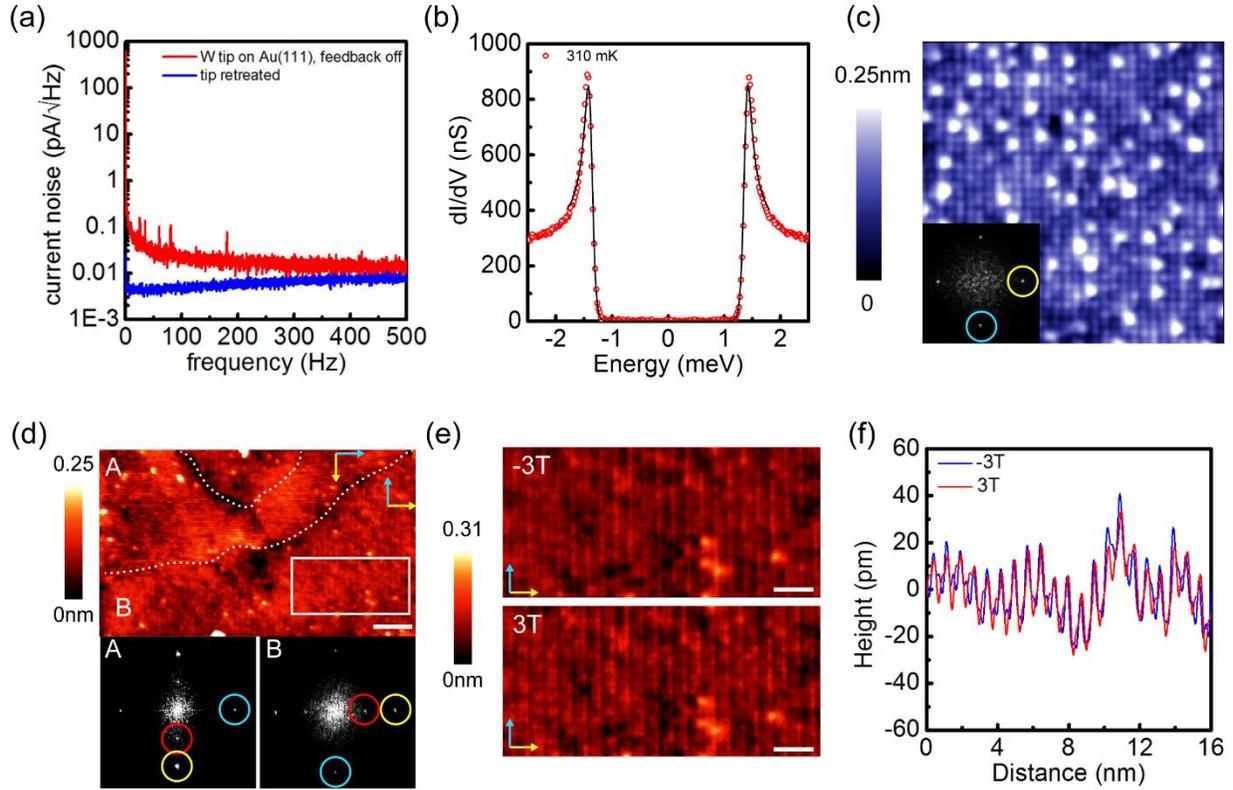

**Fig. 4.** (a) The noise spectrum of tunneling current. The blue curve represents the noise as out of tunneling. The red curve is the noise spectrum taken as W tip on the Au (111) with bias -0.1V and setpoint 600 pA at T = 300 mK. (b) The superconducting gap of Nb tip at 310 mK. The black line represents the BCS fit, which yields the effective temperature is 400 mK. (bias = 5 mV, current = 1.2 nA and lock-in modulation = 10 µV). (c) The topography of $Fe_{1+x}Te$ taken by a W tip. The FOV is $11 \times 11$ nm$^2$ with bias = 5 mV and current setpoint = 90 pA. The inset shows the Fourier transform of this image, which exhibits only the Bragg peaks (yellow and cyan circles). (d) The topography of $Fe_{1+x}Te$ acquired by a magnetic tip shows three magnetic domains. The white dashed lines represent the domain boundaries and the scale bar represents 5 nm. The yellow and cyan arrow indicate the orthorhombic a- and b-axis in each domain. The inserts show the Fourier transform of the domain A and B in (d) with the characteristic peaks for the stripe magnetic structure in $Fe_{1+x}Te$ rotate across the domain boundary. The yellow, cyan and red circles represent the charge density wave superimpose the Bragg peak, the Bragg peak and the peak due to the antiferromagnetic order, respectively. (bias = -10 mV and current setpoint = 150 pA) (e) The topography of $Fe_{1+x}Te$ taken with the same tip at the magnetic field of -3T (top) and +3T (bottom) in the same FOV, marked by the white rectangle in (d). (the scale bar = 2nm, bias = -20 mV and current = 100 pA) (f) The line profiles of topographic images in (e) averaged along the stripe direction (vertical) do not switch under magnetic field, indicating the tip is antiferromagnetic.

**Appendix**

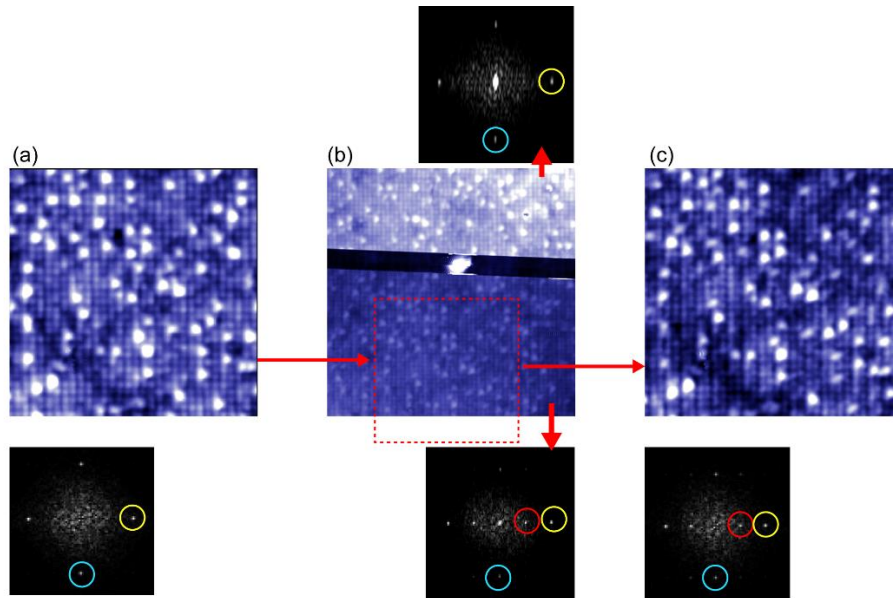

Sequence of topographic images and their corresponding Fourier transformed images show the process of the preparation of magnetic tip on the $Fe_{1+y}Te$ surface. (a) Before the preparation, (b) during the preparation (the large bright cluster is the location when the tip change occurs) and (c) after the tip preparation. The antiferromagnetic order at ($\pm 0.5$, 0) along a-axis becomes visible in the Fourier analysis after the tip change in (b). The red dashed square indicates the FOV for (a) and (c), which is the same as Fig. 4(c).